\documentclass[twocolumn]{aastex63}
\usepackage{times}
\usepackage{amsmath}
\usepackage{textcomp}
\usepackage{amssymb}
\usepackage{natbib}
\usepackage{float}
\usepackage{color}
\usepackage{graphicx}

\newcommand{\msun}{\ensuremath{M_{\odot}}}

\newcommand{\gm}{$\gamma$}

\received{\today}
\revised{\today}
\accepted{\today}
\submitjournal{ApJ}

\shorttitle{Gamma-ray Emitting GRGs}
\shortauthors{Paliya et al.}

\begin{document}
\title{Radio Morphology of Gamma-ray Sources - II. Giant Radio Galaxies}

\correspondingauthor{Vaidehi S. Paliya}
\email{vaidehi.s.paliya@gmail.com}

\author[0000-0001-7774-5308]{Vaidehi S. Paliya}
\affiliation{Inter-University Centre for Astronomy and Astrophysics (IUCAA), SPPU Campus, Pune 411007, India}
\author[0000-0002-4464-8023]{D. J. Saikia}
\affiliation{Inter-University Centre for Astronomy and Astrophysics (IUCAA), SPPU Campus, Pune 411007, India}
\author[0000-0002-5182-6289]{G. Bruni}
\affiliation{INAF $-$ Istituto di Astrofisica e Planetologia Spaziali, Via Fosso del Cavaliere 100, 00133 Rome, Italy}
\author[0000-0002-3433-4610]{Alberto Dom{\'{\i}}nguez}
\affiliation{IPARCOS and Department of EMFTEL, Universidad Complutense de Madrid, E-28040 Madrid, Spain}
\author[0000-0002-4464-8023]{C. S. Stalin}
\affiliation{Indian Institute of Astrophysics, Block II, Koramangala, Bengaluru 560034, Karnataka, India}

\begin{abstract}
Giant radio sources, including galaxies and quasars (hereafter GRGs), are active galactic nuclei (AGN) hosting relativistic jets with source sizes exceeding the projected length of 0.7 Mpc. They are crucial to understanding the evolution of radio sources and their interaction with the surrounding environment. Some of these enigmatic objects, e.g., NGC~315, have also been reported as \gm-ray emitters. Since GRGs are thought to be aligned close to the plane of the sky, they are invaluable targets to explore the radiative mechanisms responsible for the observed \gm-ray emission. We have carried out a systematic search of \gm-ray emitting GRGs using sensitive low-resolution radio surveys, such as by Low Frequency Array, NRAO VLA Sky Survey, and Rapid ASKAP Continuum Survey, and considering the fourth data release of the fourth Fermi-Large Area Telescope \gm-ray source (4FGL-DR4) catalog. By carefully inspecting the radio maps of all AGN included in the 4FGL-DR4 catalog, we have identified 16 \gm-ray emitting GRGs, including 8 of them being reported as GRGs for the first time. Some of their observed parameters, e.g., core dominance, appeared to differ from that found for the non-\gm-ray detected GRG population, possibly due to the relatively small viewing angle of the \gm-ray emitting jet. The observed \gm-ray properties of these objects were found to be similar to non-GRG \gm-ray emitting misaligned AGN. We conclude that the origin of the \gm-ray emission could be similar in both source populations.

\end{abstract}


\section{Introduction}
Relativistic jets manifest some of the most intriguing processes involving the interaction of matter and energy at the centers of galaxies. The projected lengths of the sources fed by these jets range from a few parsecs (pc), for compact symmetric objects, to $\sim$Mpc scales for giant radio sources \citep[e.g.,][]{1994ApJ...432L..87W,1974Natur.250..625W,2024Natur.633..537O}. Active galactic nuclei (AGN) hosting powerful relativistic jets are often termed radio-loud AGN. They have been classified as Fanaroff-Riley Type I (FR~I) and FR~II radio sources based on their radio power, with the latter being the more luminous ones \citep[][]{1974MNRAS.167P..31F}. Originally, radio sources were classified in this scheme based on their distinct radio morphologies, with FR~Is exhibiting edge-darkened, diffuse radio structures. The FR~IIs, on the other hand, showed large-scale, edge-brightened radio lobes with bright hotspots. However, recent Low-Frequency Array (LOFAR) observations have revealed several outliers, e.g., low-power 
FR~IIs and vice versa \citep[][]{2019MNRAS.488.2701M}. The radio-loud AGN have also been classified as low-excitation radio galaxies (LERGs) and high-excitation radio galaxies (LERGs) based on the accretion activity. The HERGs exhibit radiatively efficient accretion \citep[$L_{\rm bol.}/L_{\rm Edd}>0.01$; e.g.,][]{2012MNRAS.421.1569B}. The relationship between the FR classes and mode of accretion, i.e., radiatively efficient/inefficient, is complex with no clear pattern emerging from observations \citep[][]{2019MNRAS.488.2701M,2022MNRAS.511.3250M}. 

Among the jetted AGN population, giant radio sources, associated both with radio galaxies and quasars (hereafter GRGs), are one of the most intriguing members exhibiting gigantic source sizes produced by the relativistic jets \citep[projected length $\gtrsim$0.7 Mpc;][]{1974Natur.250..625W}. For example, recent LOFAR observations have revealed the source {\it Porphyrion} hosting a jet exceeding 7 Mpc \citep[][]{2024Natur.633..537O}. The largest catalog of GRGs has recently been published by \citet[][]{2024A&A...691A.185M} reporting more than 11000 sources \citep[see also,][for a review]{2023JApA...44...13D}. A majority of GRGs are reported to be of FR~II type, and only a minor fraction has been identified as FR~Is \citep[cf.][]{1999MNRAS.309..100I,2018ApJS..238....9K,2020A&A...635A...5D,2020A&A...642A.153D}. Some GRGs also exhibit signatures of episodic jet activity \citep[e.g.,][]{2000MNRAS.315..371S,2017MNRAS.471.3806K,2019A&A...622A..13M,2023MNRAS.525L..87C,2024arXiv240813607D}. These objects are usually powered by massive black holes ($>10^8$ \msun) similar to other non-GRG radio galaxies and quasars \citep[cf.][]{2012MNRAS.426..851K,2020A&A...642A.153D}. A major fraction of these objects exhibit an LERG-type accretion activity \citep[][]{2012MNRAS.426..851K}.

\citet[][]{2018MNRAS.481.4250U} carried out a broadband X-ray spectral analysis of a sample of hard X-ray selected GRGs and reported the X-ray emission to originate from the X-ray corona illuminated by radiatively efficient accretion. The nuclear X-ray emission was more luminous than the jet power and radio lobe luminosity, indicating a restarted AGN activity. Subsequent high-resolution radio observations have revealed the presence of young radio sources in the cores of hard X-ray selected GRGs, thus indicating possible rejuvenated AGN activity episode \citep[][]{2019ApJ...875...88B,2020MNRAS.494..902B}. 

The deepest survey of the \gm-ray sky is being conducted by the Fermi Large Area Telescope \citep[LAT;][]{2009ApJ...697.1071A}. In its first 14 years of operation, summarized in the fourth data release of the fourth Fermi-LAT $\gamma$-ray source catalog (4FGL-DR4), the Fermi-LAT has detected 7194 $\gamma$-ray sources in the 50 MeV to 1 TeV range \citep[][]{2022ApJS..260...53A,2023arXiv230712546B}, the majority of which are blazars, i.e. AGN with jets directed close to our line of sight. In contrast, only a small number of misaligned AGN ($\sim$50) were reported and they have been found to be contributing up to $\sim$10\% of the extragalactic \gm-ray background \citep[][]{2022ApJ...931..138F}. The 4FGL-DR4 catalog also reported 1624 blazar candidates of uncertain type, i.e., sources whose multi-wavelength behavior are similar to blazars though they lack optical spectroscopic classification \citep[][]{2011ApJ...743..171A,2022ApJS..260...53A}. All in all, there are 4069 \gm-ray emitting AGN present in the 4FGL-DR4 catalog. Furthermore, the \gm-ray emission from misaligned AGN has been explained with the synchrotron self Compton process along with more complex models, such as spine-sheath structured jets \citep[][]{2018ApJ...855...93F,2010MNRAS.402.1649G,2011A&A...533A..72M}.

The properties of GRGs in the \gm-ray band are not well explored. Only a few objects of this class have been studied, that too as a member of the general \gm-ray detected misaligned jetted AGN population \citep[][]{2022ApJ...931..138F}. Considering the AGN unification, GRGs are seen at large viewing angles. Indeed, the core dominance fraction, an indicator of the jet viewing angle, was found to decrease with increasing projected jet length \citep[e.g.,][]{1982JApA....3..465K}. The GRGs also have smaller core dominance compared to non-GRG radio sources \citep[cf.][]{2022A&A...660A..59M}. This observation can explain the paucity of \gm-ray detected GRGs in the \gm-ray source catalogs since \gm-ray emission is highly sensitive to the viewing angle \citep[e.g.,][]{1995ApJ...446L..63D}. Interestingly, \citet[][]{2024ApJ...965..163Y} reported the detection of \gm-ray emission from the radio lobes of a GRG NGC 6251 ($z=0.02$), indicating a different \gm-ray emitting region location or physical mechanism compared to that usually considered for beamed AGN or blazar population \citep[cf.][]{2017ApJ...851...33P}. 

In \citet[][hereafter Paper I]{2024ApJ...976..120P}, we started a program to identify \gm-ray emitting misaligned jetted AGN population by utilizing the sensitive radio data provided by ongoing surveys, e.g., LOFAR. We used small beamsize survey data, e.g., Very Large Array Sky Survey (VLASS), to identify radio sources that might have remained unresolved in earlier surveys and found 149 double-lobed objects exhibiting multi-wavelength behavior, e.g., radio spectrum and optical spectroscopic properties, similar to misaligned AGN \citep[see also,][]{2010ApJ...720..912A,2020JHEAp..27...77C}. However, a drawback of choosing high-resolution data is that some misaligned sources with large-scale diffuse radio emissions, such as GRGs, may have been missed. Therefore, in this work, we adopted radio surveys with the lowest resolution, thus the largest restoring beams, to identify potential \gm-ray emitting GRGs and improve the completeness of the \gm-ray detected misaligned AGN population. Section~\ref{sec2} briefly describes the sample selection and radio surveys whose data were adopted in this work. We present our findings in Section~\ref{sec3} and discuss them in Section~\ref{sec4}. We summarize the results in Section~\ref{sec6}. Throughout, we have adopted the flux density $F_{\nu}\propto\nu^{\alpha}$, where $\alpha$ is the radio spectral index. A flat cosmology with $H_0 = 70~{\rm km~s^{-1}~Mpc^{-1}}$ and $\Omega_{\rm M} = 0.3$ was used.

\section{Sample Selection and Multi-wavelength Catalog}\label{sec2}
\subsection{The Sample}
We used the 4FGL-DR4 catalog which lists 7194 \gm-ray sources \citep[][]{2022ApJS..260...53A,2023arXiv230712546B}. From this sample, we considered 4069 objects whose multiwavelength counterparts have been reported to be AGN.

\subsection{Radio Catalogs}

\begin{deluxetable*}{ccccccccccc}
\tabletypesize{\small}
\tablecaption{The list of \gm-ray emitting misaligned jetted AGN hosting Mpc-scale radio sources.\label{tab:basic_info}}
\tablewidth{0pt}
\tablehead{
\colhead{4FGL Name} & \colhead{Other name} & \colhead{$z$} & \colhead{LAS} & \colhead{$F_{\rm core}$} & \colhead{$F_{\rm total}$} & \colhead{$\alpha$} & \colhead{$C_{\rm r}D$} & \colhead{Survey} &  \colhead{Morph.} & \colhead{Ref.}\\
\colhead{[1]} & \colhead{[2]} & \colhead{[3]} & \colhead{[4]} & \colhead{[5]} & \colhead{[6]} & \colhead{[7]} & \colhead{[8]} & \colhead{[9]} & \colhead{[10]} & \colhead{[11]}}
\startdata
  J0037.9+2612 &   GB6 B0034+2556 &  0.148 &    410 (1.07)  &   80   &  550  &  -0.37 & 0.32 & L,V,L  &FRI & A20\\
  J0049.0+2252 &   PKS J0049+2253 &  0.264 &    220 (0.89)  &   141   &  1660  &  -0.51 & 0.10 & L,V,L  &FRI & A20\\
  J0057.7+3023 &   NGC 315        &  0.017 &    5300 (1.84)  &   500   &  13400  &  -0.30 & $-$0.35 & L,V,L  &FRI-II & J00\\
  {\bf J0126.5$-$1553} &   PMN J0127$-$1556 &  0.988 &    94 (0.76)  &   25   & 150 &  -0.55 & $-$0.03& R,V,R  &FRII & F22 \\
  J0312.9+4119 &   B3 0309+411B   &  0.134 &    530 (1.27)  &   310   &  520  &  -0.01 & 0.49 & N,V,N  &FRI & K22\\
  {\bf J0525.6$-$2008} &   PMN J0525$-$2010 &  0.092 &    700 (1.20)  &   110  &  455  &  -0.19 & $-$0.04 & R,V,R  &FRI & P21\\
  {\bf J0617.7$-$1715} &   IVS B0615$-$172  &  0.098 &    650 (1.18)  &   150   &  805  &  -0.22 & $-$0.33 & N,V,N  &FRI & S13 \\
  {\bf J0946.0+4735} &   RX J0946.0+4735&  0.569 &    210 (1.41)  &   15   &  66  &  -0.32 & 0.71 & L,V,L  &FRI & A20\\
  J1202.4+4442 &   GB6 J1202+4444 &  0.297 &    180 (0.79)  &   45  & 390  &  -0.46 & 0.27 & L,V,L  &FRI & A20\\
  J1226.9+6405 &   GB6 J1226+6406 &  0.110 &    425 (0.86)  &   40   &  650  &  -0.18 & $-$0.06 & L,V,L  &FRI-I & A20\\
  {\bf J1230.9+3711} &   GB6 J1231+3711 &  0.218 &    255 (0.91)  &   17   &  190  &  -0.74 & $-$0.67 & N,V,N  &FRI & A20\\
  J1630.6+8234 &   NGC 6251       &  0.020 &    3400 (1.38)  &   510   &  2150  &  -0.05 & $-$0.23 & N,V,N  &FRI-II & W03\\
  {\bf J2228.5+2211} & GB6 B2226+2156 &  0.710 &  190 (1.39)  &   30 &  120 &  -0.27 & 0.78 & L,V,L  &FRI & A20\\
  {\bf J2253.3+3233} & TXS 2250+323   &  0.258 & 250 (1.00)  &   125   &  340 &  -0.05 & 0.89 & L,V,L  &FRI & T17\\
  {\bf J2302.8$-$1841} &   PKS 2300$-$18    &  0.128 &    325 (0.75)  &   235   &  1981  &  -0.55 & $-$0.41 & R,V,R  &FRI & J09 \\
  J2333.9$-$2343 &   PKS 2331$-$240   &  0.047 &    1130 (1.05)  &   970   &  1150 &   0.13 & 1.01 & N,V,N  &FRII & K22\\
\enddata
\tablecomments{The column details are as follows: (1) 4FGL name; (2) counterpart name; (3) redshift; (4) largest angular size (and the corresponding linear size) in arcsec (in Mpc); (5) and (6) core and total flux densities, respectively (in mJy); (7) radio spectral index provided in SPECFIND; (8) core dominance; (9) the surveys which were used to estimate the LAS, core and total flux densities, in respective order; V=VLASS, L=LOFAR, N=NVSS, and R=RACS; and (10) morphological classification. In the first column, source names written in boldface are reported as GRGs for the first time in this work. The last column provides the references of the optical spectra and redshift: A20: \citet[][]{2020ApJS..249....3A}, J00: \citet[][]{2000ApJS..126..331J}, F22: \citet[][]{2022Univ....8..587F}, K22: \citet[][]{2022ApJS..261....2K}, P21: \citet[][]{2021MNRAS.504.3338P}, S13: \citet[][]{2013AJ....146..127S}, W03: \citet[][]{2003AJ....126.2268W}, T17: \citet[][]{2017AJ....153..157T}, and J09: \citet[][]{2009MNRAS.399..683J}.}
\end{deluxetable*}

{\it LOFAR Two-metre Sky Survey}: The second data release of the LOFAR Two-metre Sky Survey (LOTSS-DR2) provides the cutout images at 144 MHz with resolutions of 6$^{\prime\prime}$ and 20$^{\prime\prime}$ \citep[][]{2022A&A...659A...1S}. Since our primary objective was to identify the large-scale diffuse radio emission, we downloaded $30^{\prime}\times 30^{\prime}$ cutout images of 935 objects that have a resolution of 20$^{\prime\prime}$\footnote{\url{https://lofar-surveys.org/dr2\_release.html}}.

{\it NRAO-VLA Sky Survey or NVSS}: It is a 1.4-GHz continuum survey covering the sky north of $-$40$^{\circ}$ declination. The resolution of the survey is 45$^{\prime\prime}$ and the rms brightness fluctuations of about 0.45 mJy/beam \citep[see,][for details]{1998AJ....115.1693C}. We downloaded $1\times1$ deg$^{2}$ cutout images of 4531 sources from the SkyView Virtual Observatory website\footnote{\url{https://skyview.gsfc.nasa.gov/current/cgi/query.pl}}.

{\it Rapid ASKAP Continuum Survey or RACS}: This first data release of this survey covered the whole sky between the declinations $-$80$^{\circ}$ to +30$^{\circ}$ at an operating frequency of 888 MHz and a resolution of 25$^{\prime\prime}$ \citep[][]{2021PASA...38...58H}.  We considered the $30^{\prime}\times 30^{\prime}$ cutout images of 2337 objects and downloaded them from the CSIRO ASKAP science data archive\footnote{\url{https://research.csiro.au/casda/}}.

\section{Results}\label{sec3}
\begin{figure*}
\hbox{
    \includegraphics[scale=0.5]{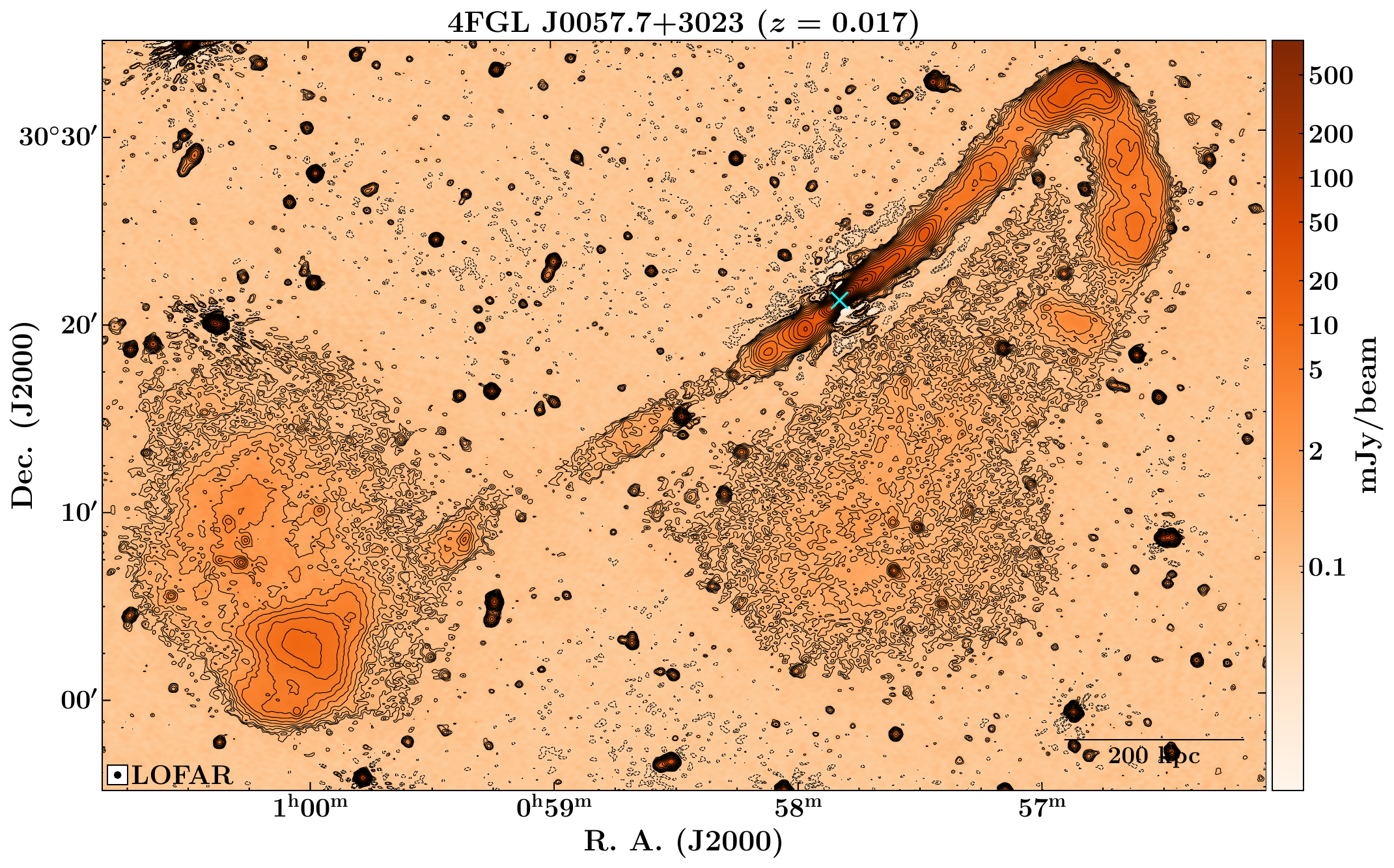}
    }
    \hbox{
    \includegraphics[scale=0.23]{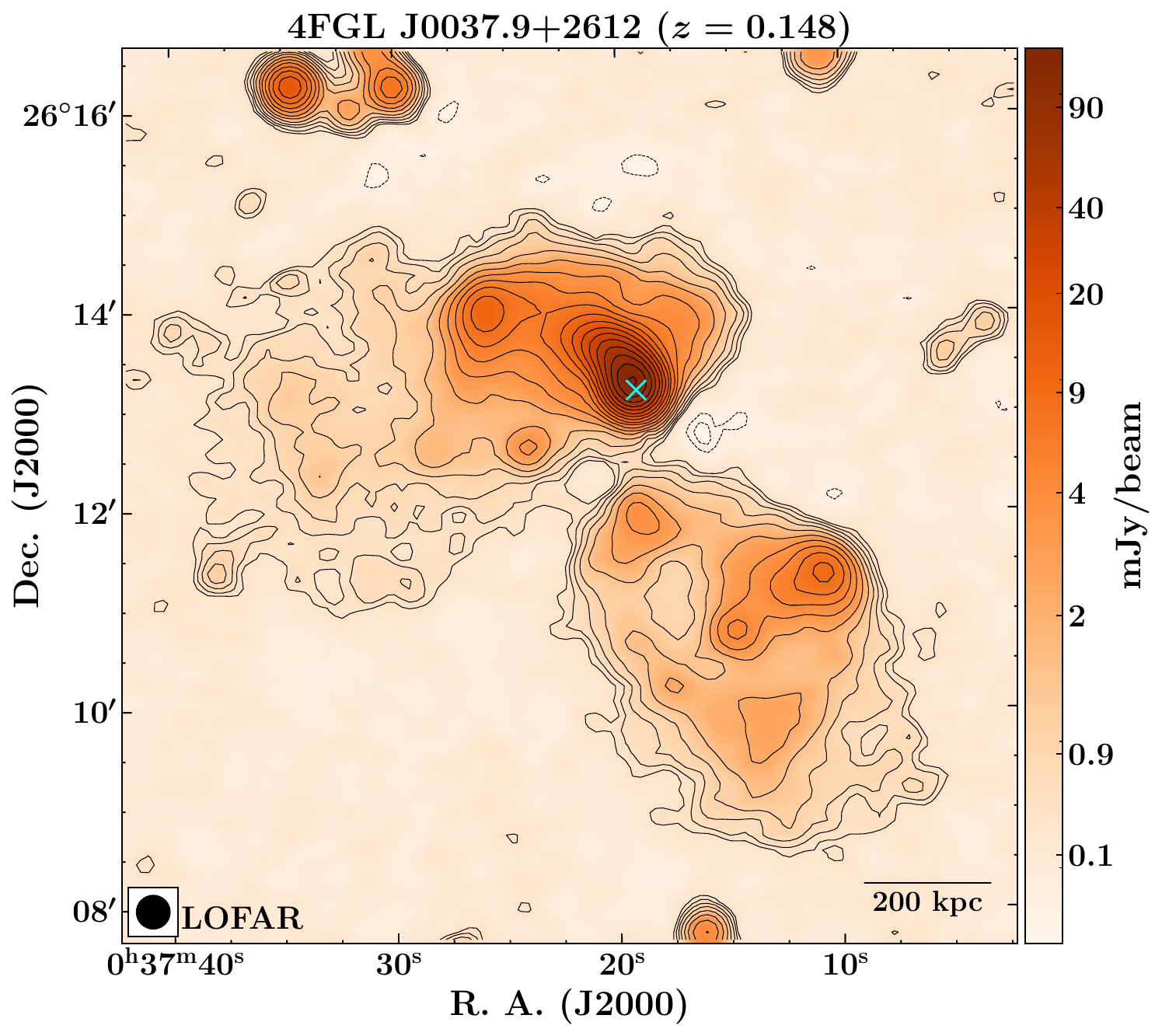}
    \includegraphics[scale=0.23]{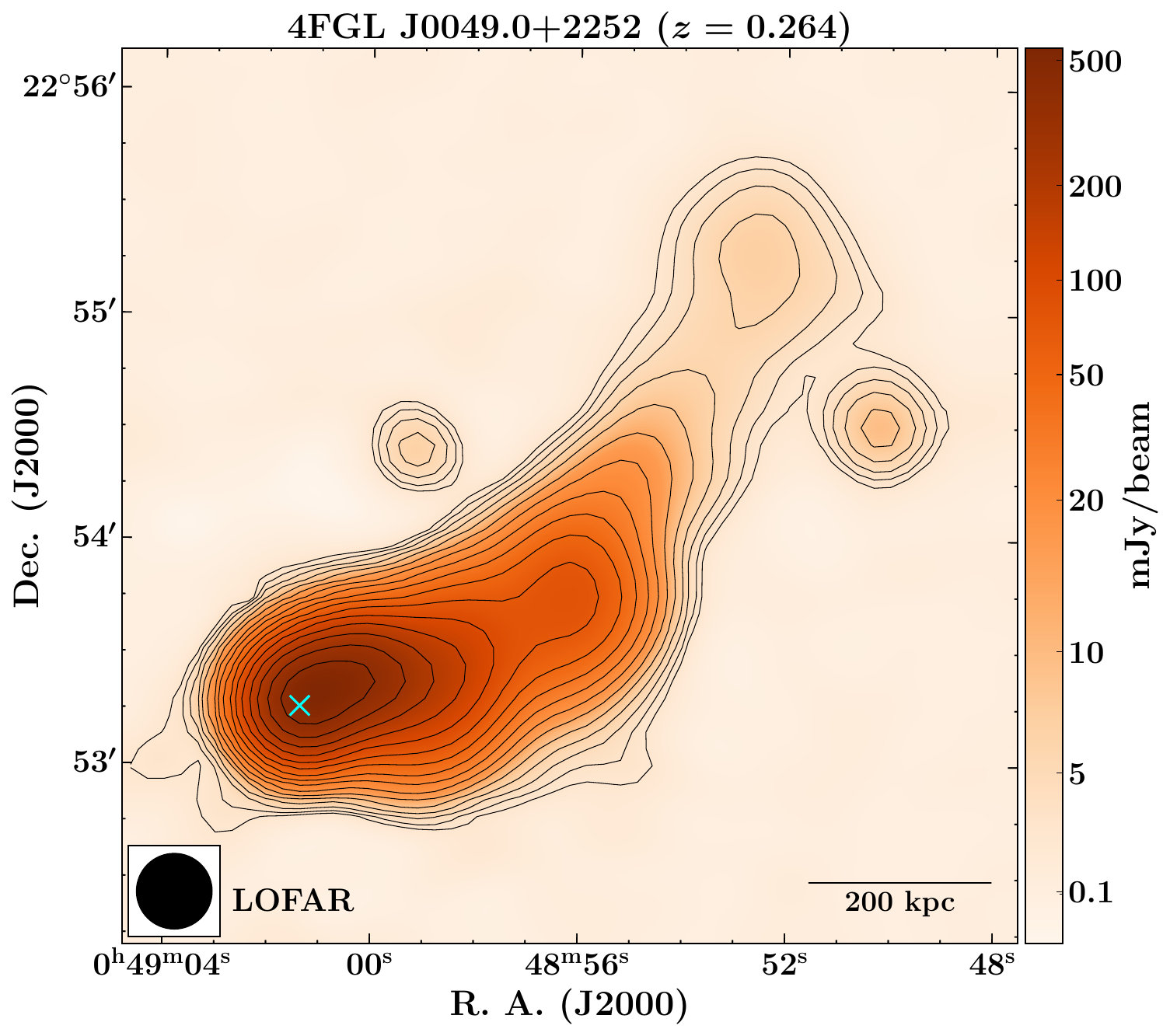}
    \includegraphics[scale=0.23]{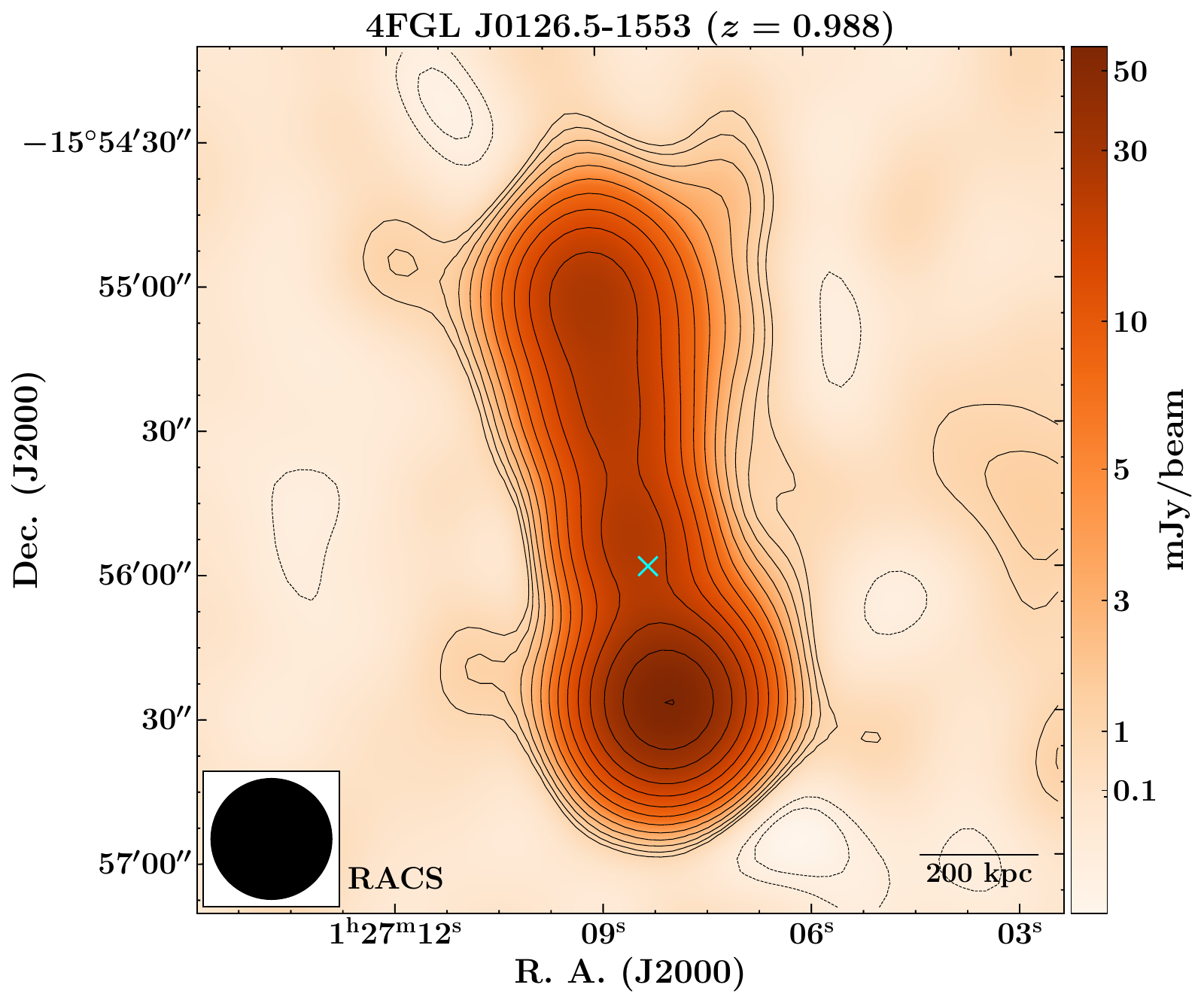}
    }
\caption{The radio morphologies of \gm-ray detected GRGs. In all plots, the contour levels start at 3$\times$rms$\times$($-$2, $-\sqrt{2}$, $-$1, 1) and increases in multiples of $\sqrt{2}$. The beam size and the name of the considered survey are shown in the bottom left corner. The north is up and east to the left. The crosses mark the positions of the optical objects.}\label{fig:vlass1}
\end{figure*}

We visually examined the radio morphology of all \gm-ray sources and retained those with projected largest angular size (LAS) $\gtrsim$1 arcminute. This is because at redshifts 0.1, 0.5, and 1, a linear size of 700 kpc corresponds to 6.3$^{\prime}$, 1.9$^{\prime}$, and 1.4$^{\prime}$, respectively. Therefore, any radio source with an LAS exceeding one arcminute will likely be a GRG candidate. We followed a strategy to compute the parameters similar to that adopted in our previous work (Paper I). In particular, to derive the LAS, we used the survey data in which the source is best resolved. The LAS was measured from the 3$\sigma$ outermost flux density contour level for FR~I morphologies. For FR~IIs, on the other hand, it was estimated from the brightest pixel on the hotspot. To determine the 3$\sigma$ contour levels, the rms noise value ($\sigma$) was computed from a nearby region free from the source contamination. Finally, only sources with available spectroscopic/photometric redshifts were considered to convert the measured LAS to physical units in kpc. This exercise led to a final sample of 16 \gm-ray emitting GRGs. Table~\ref{tab:basic_info} provides the basic information about these sources, and we show their radio maps in Figure~\ref{fig:vlass1}.
 
We also derived the core dominance, which is considered a good proxy for the orientation of the beamed emission since the lobe emission is isotropic whereas the core emission is Doppler boosted \citep[e.g.,][]{1997MNRAS.284..541M}. The following relation was adopted:

\begin{equation}
    C_{\rm r}D=\log\left(\frac{F_{\rm core}}{F_{\rm ext}}(1+z)^{\alpha_{\rm core}-\alpha_{\rm ext}} \right),
\end{equation}

\begin{figure*}
\figurenum{1}
    \hbox{
    \includegraphics[scale=0.23]{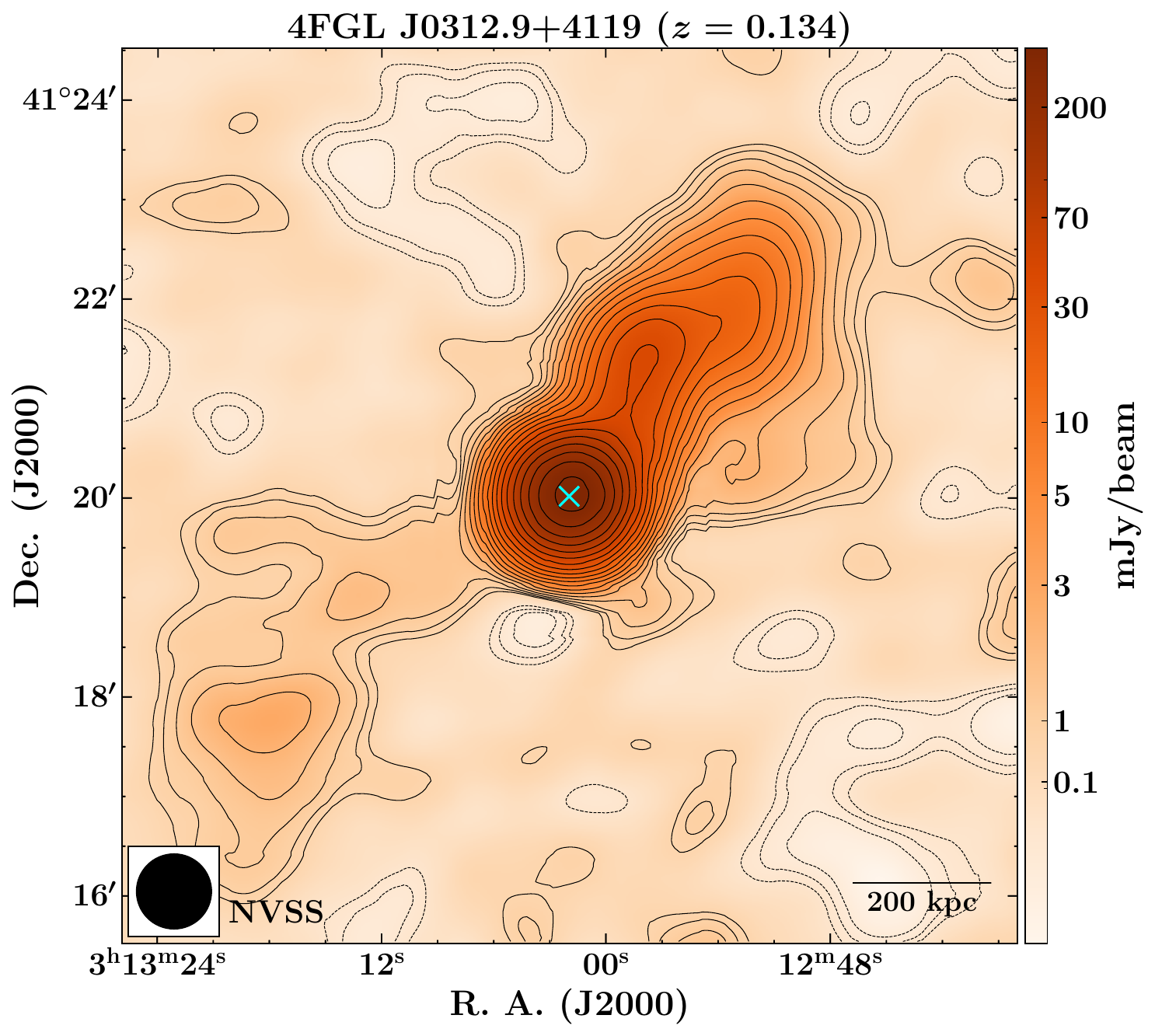}
    \includegraphics[scale=0.23]{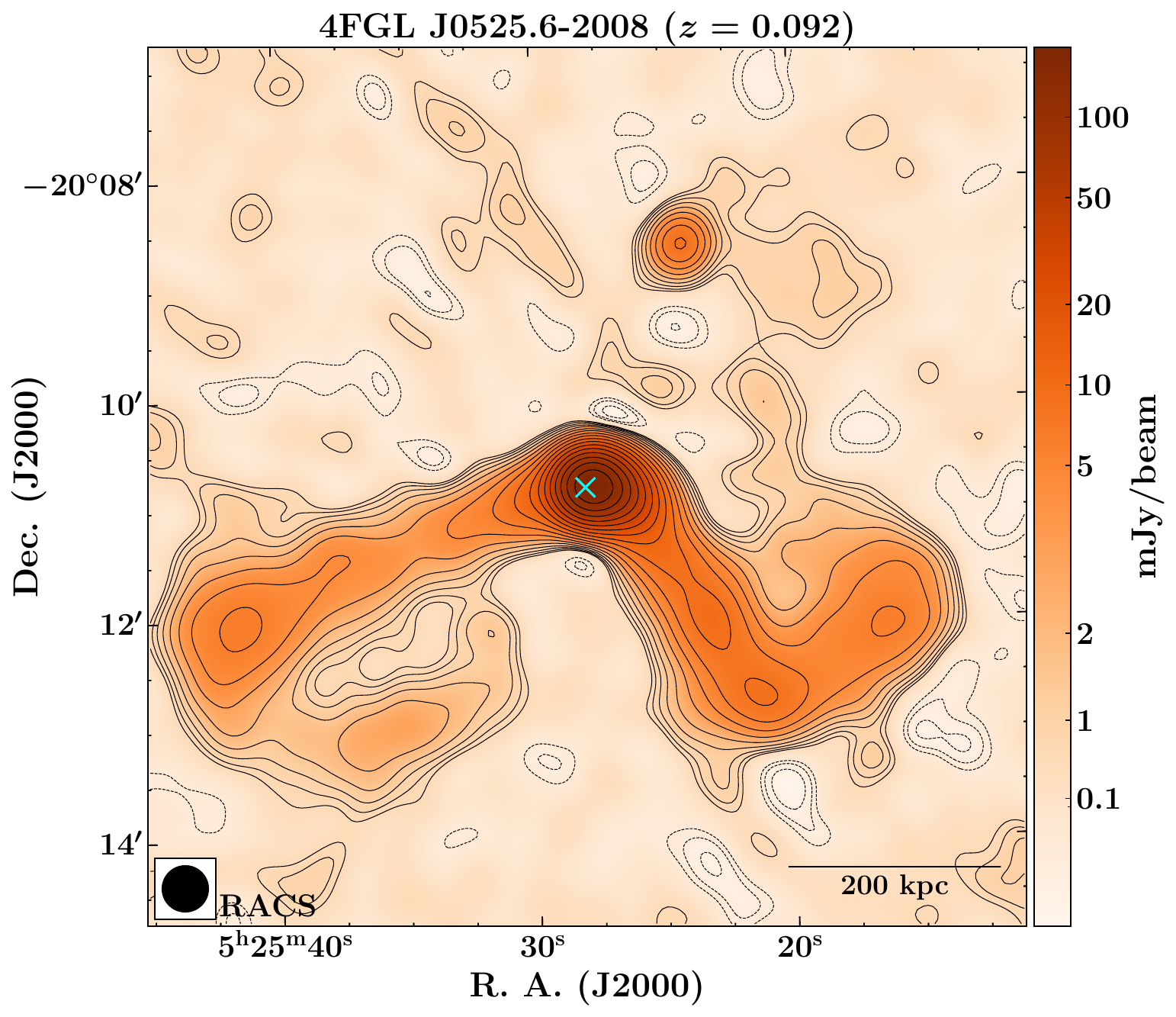}
    \includegraphics[scale=0.23]{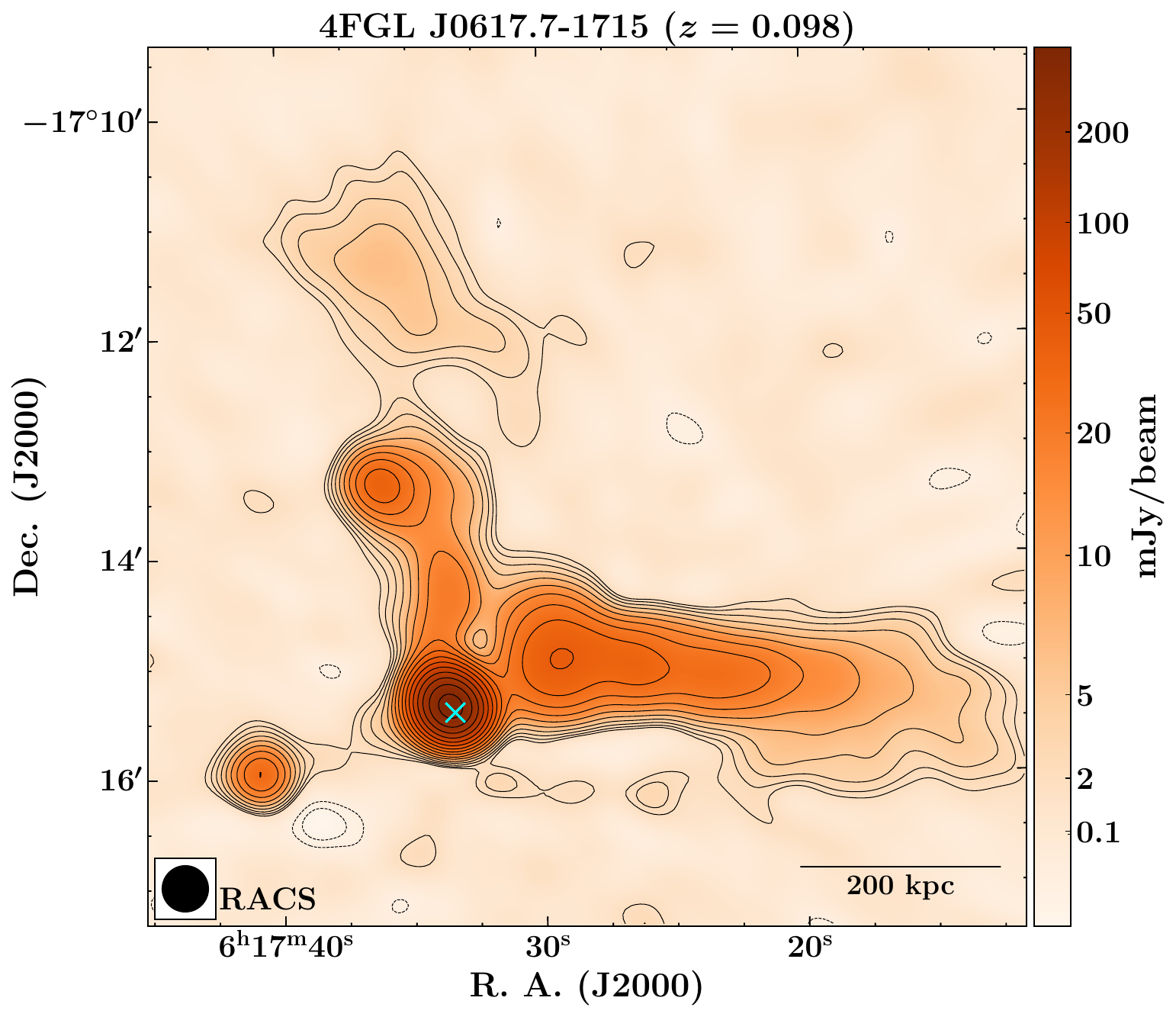}
    }
    \hbox{
    \includegraphics[scale=0.23]{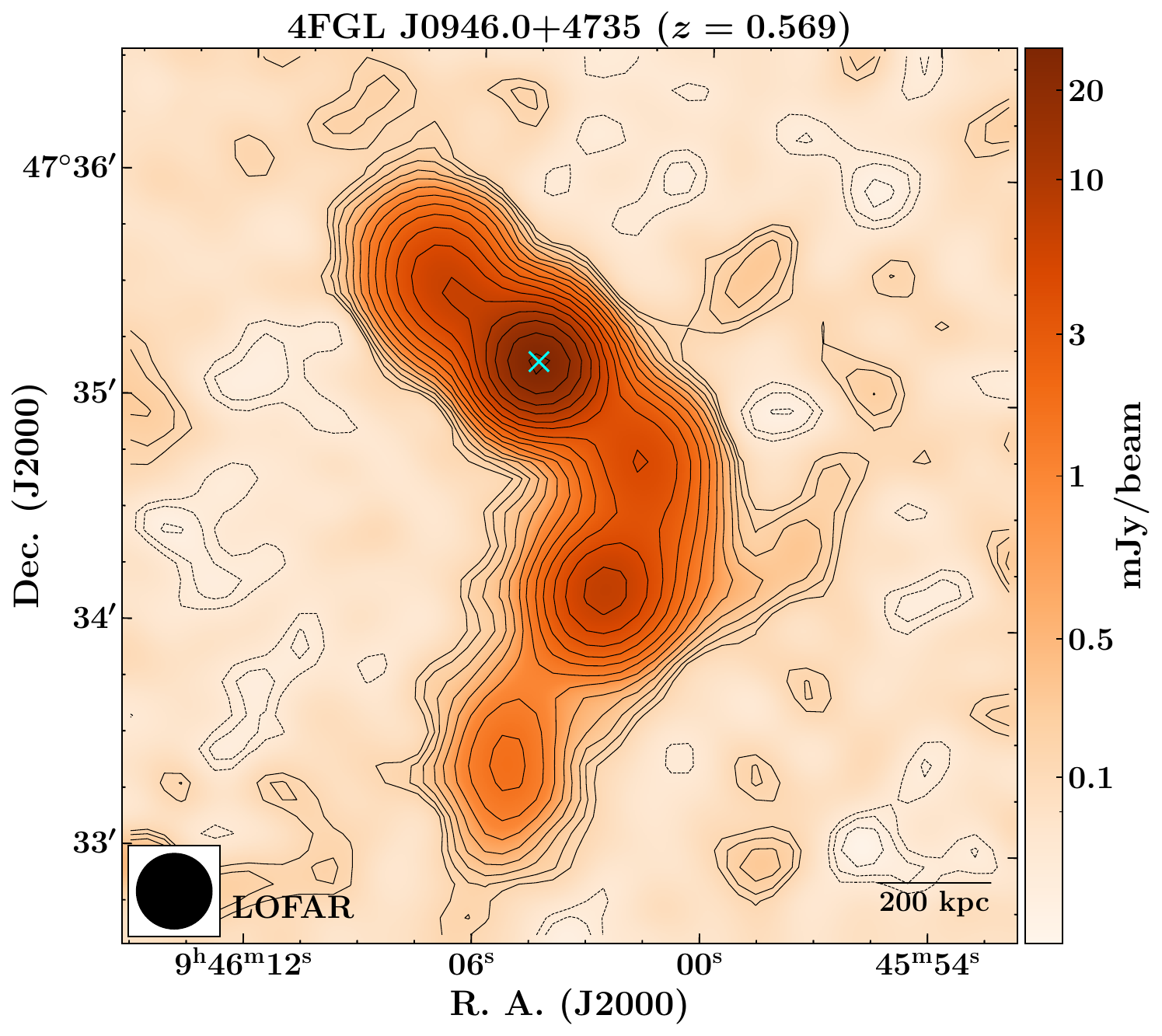}
    \includegraphics[scale=0.23]{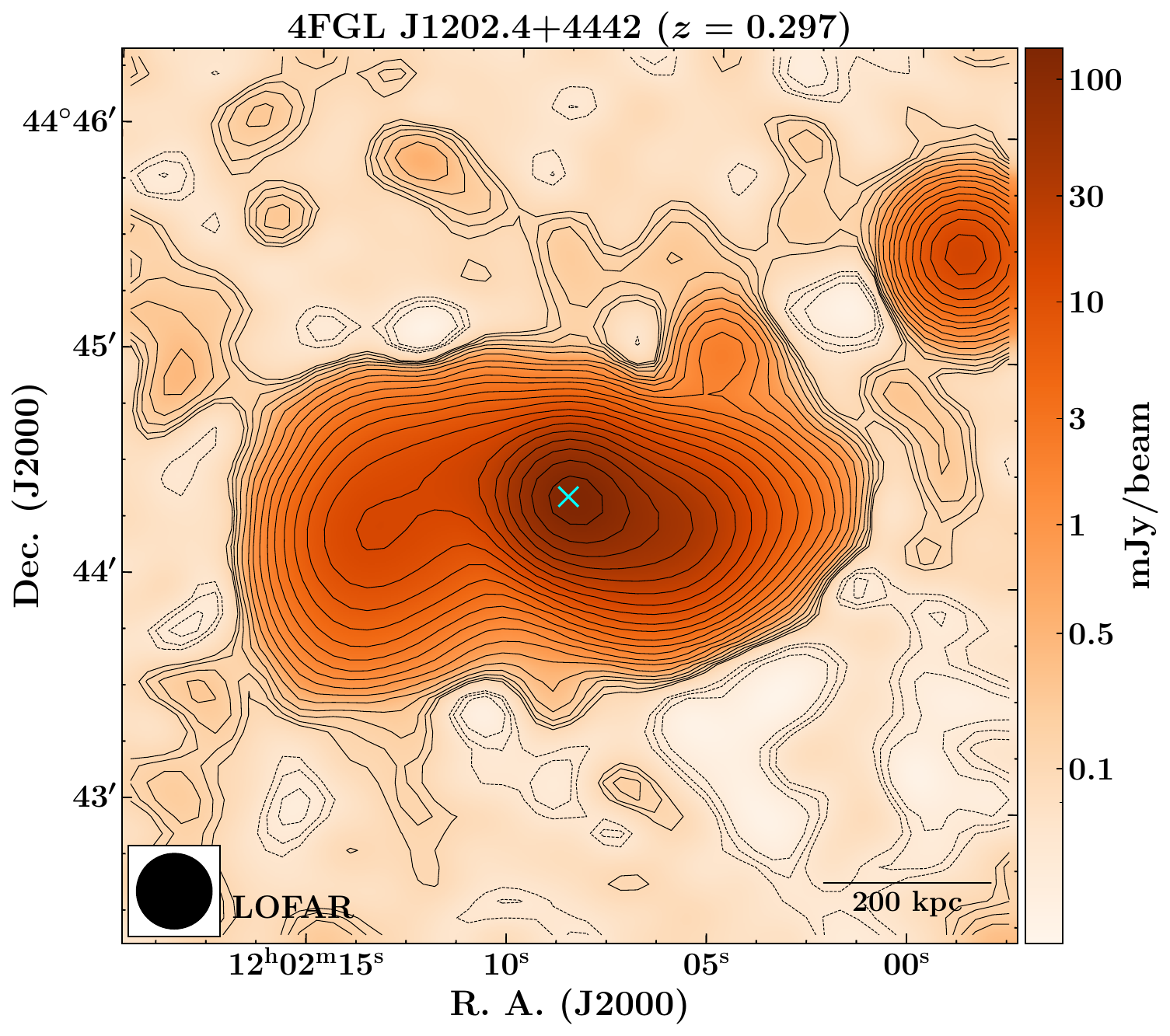}
    \includegraphics[scale=0.23]{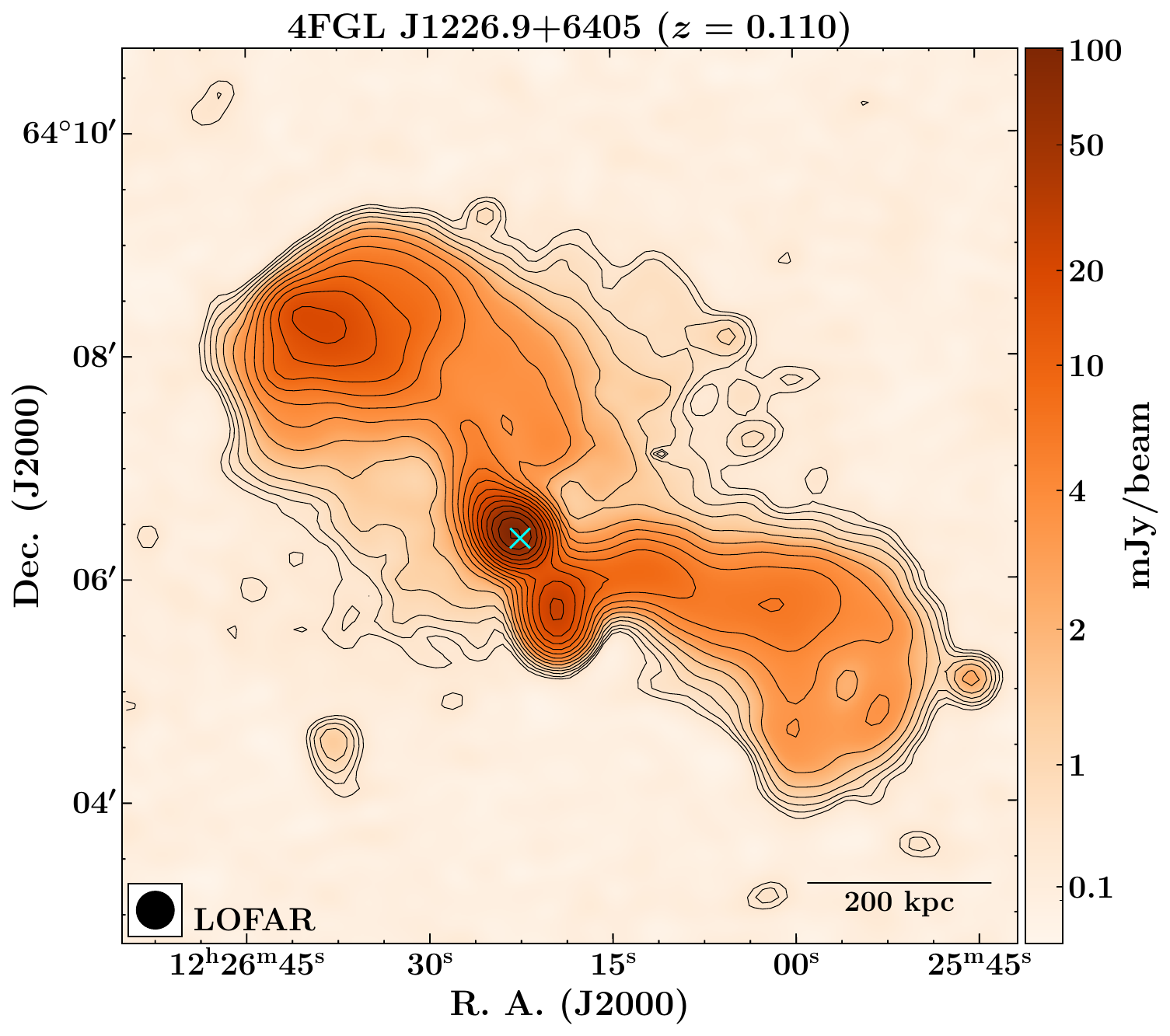}
    }
    \hbox{
    \includegraphics[scale=0.23]{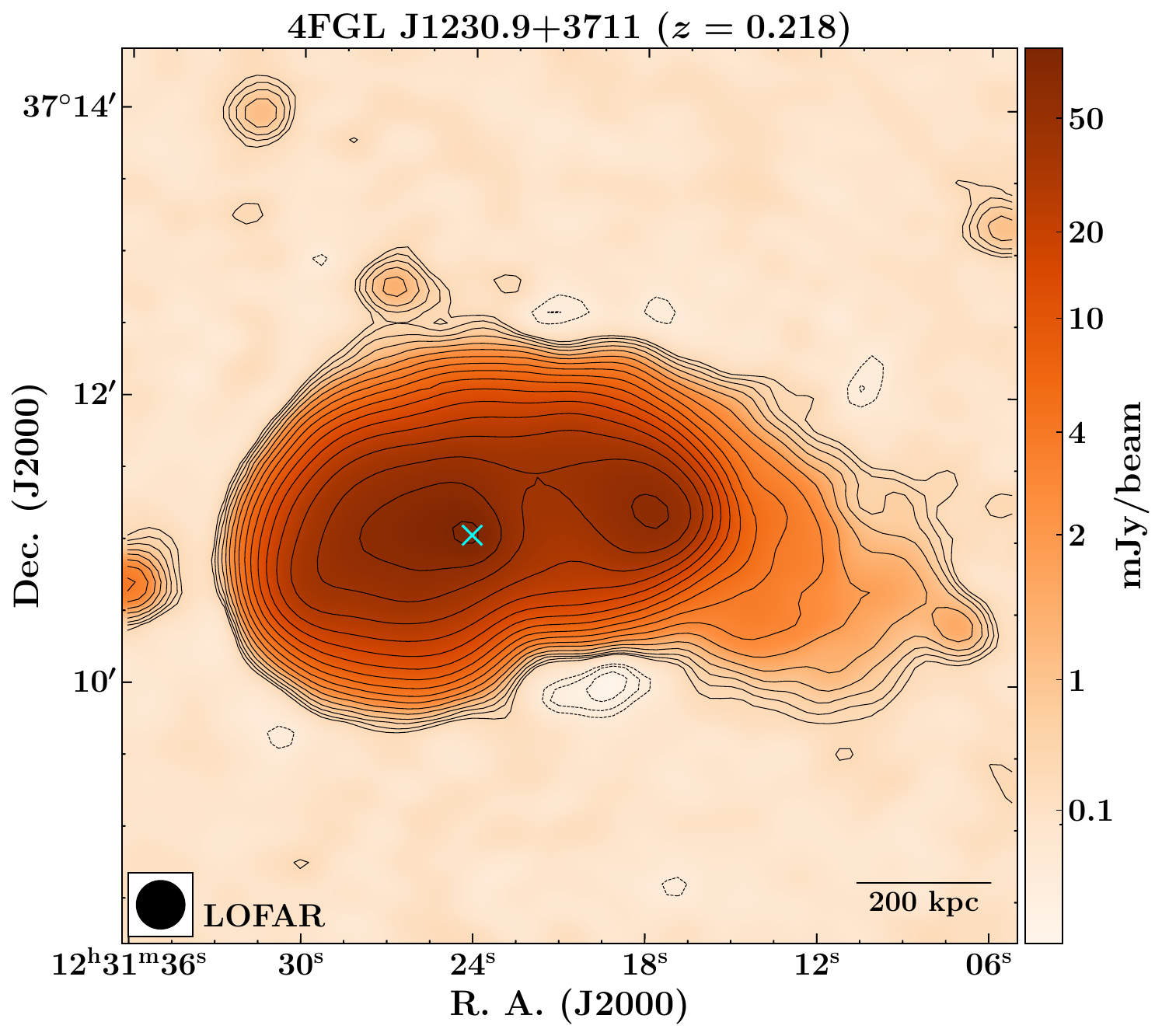}
    \includegraphics[scale=0.23]{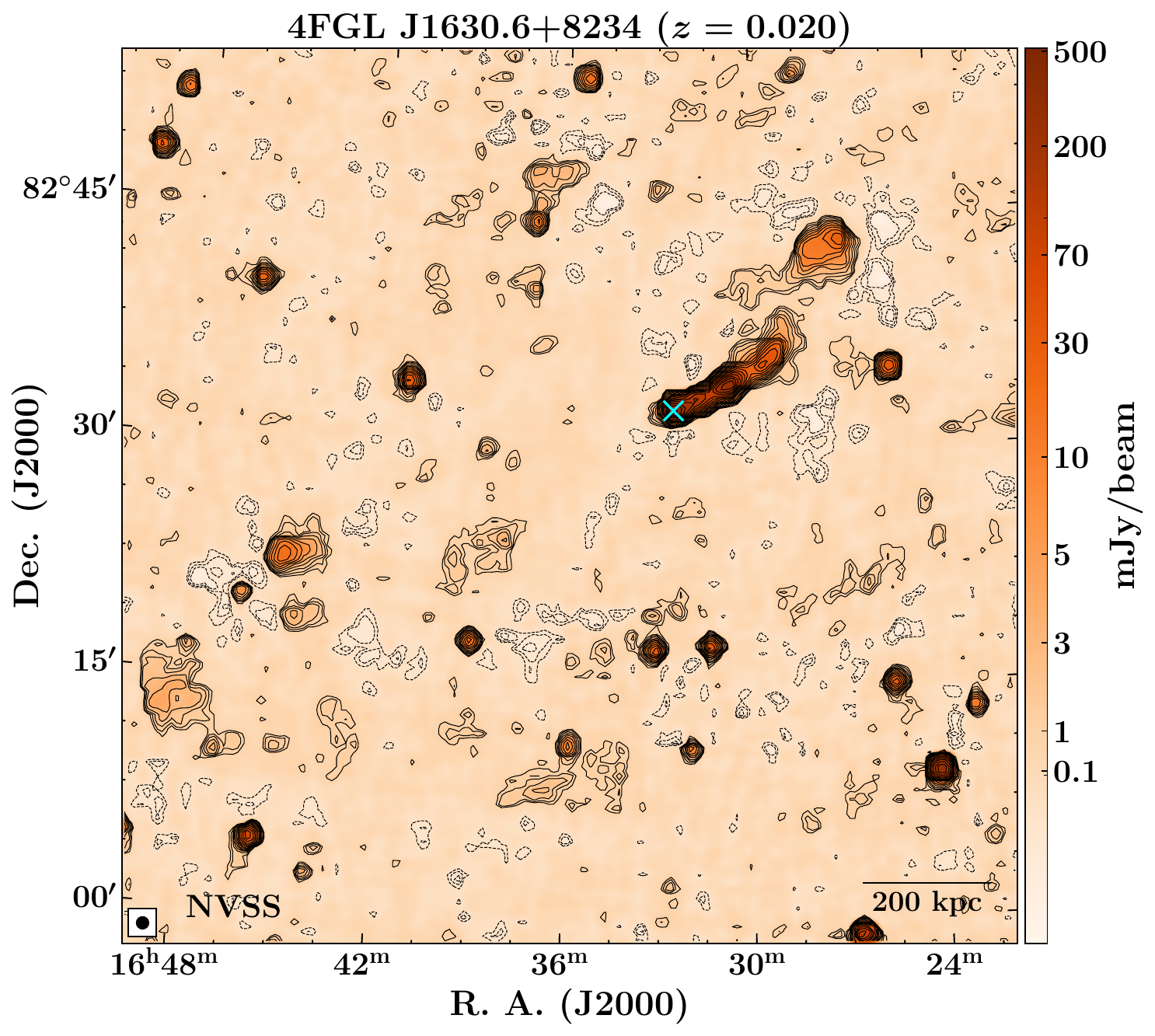}
    \includegraphics[scale=0.23]{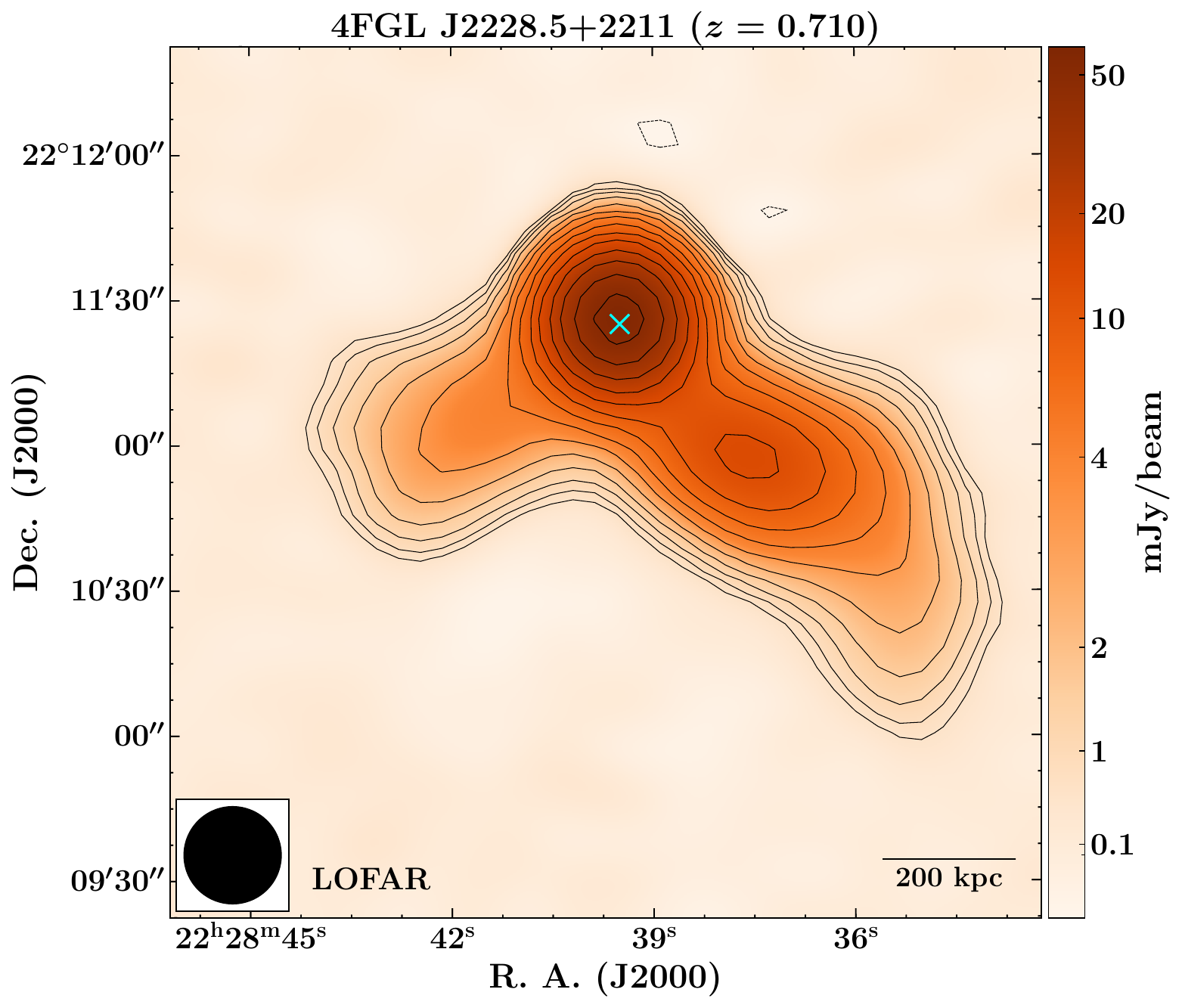}
    }
    \hbox{
    \includegraphics[scale=0.23]{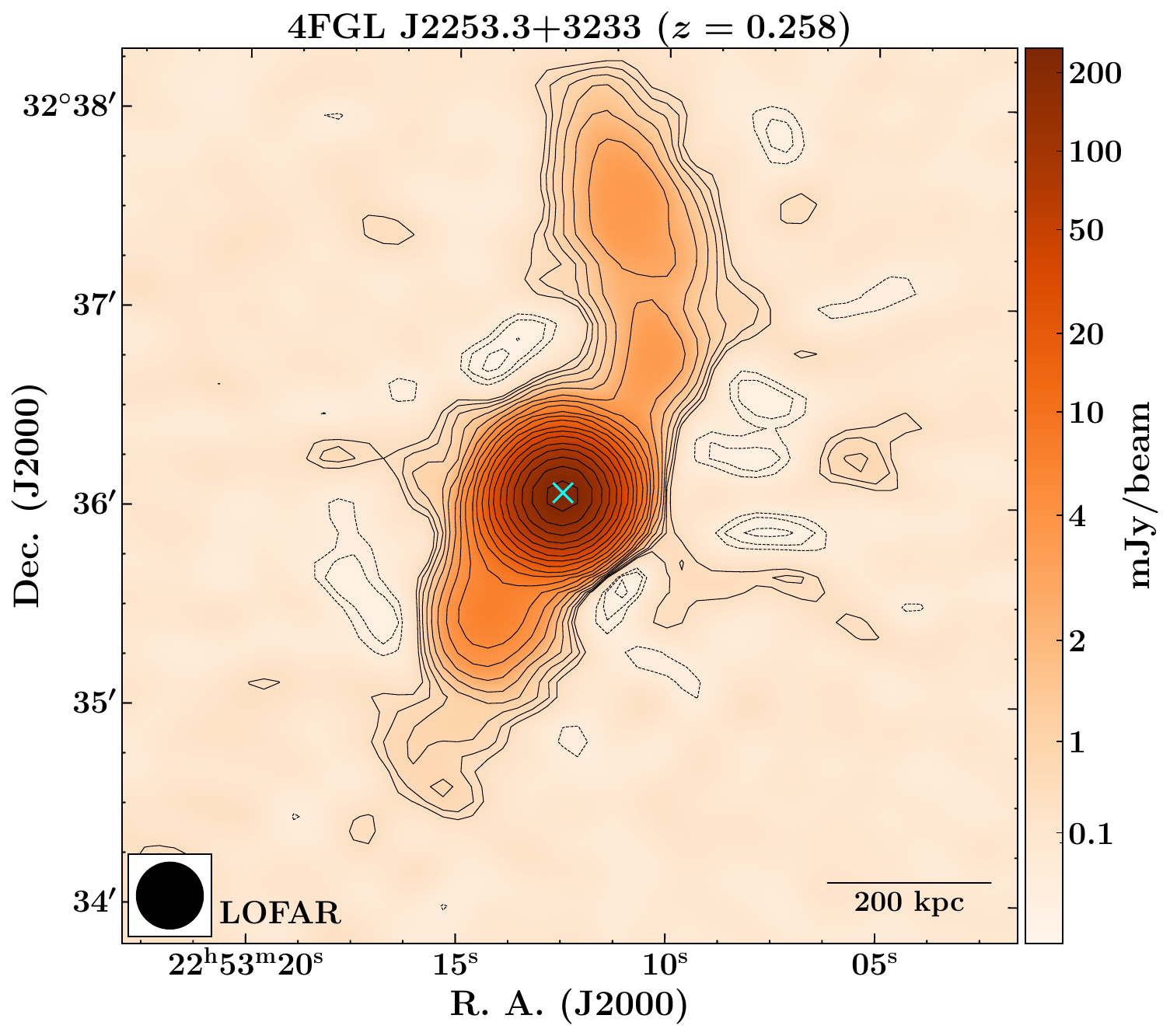}
    \includegraphics[scale=0.23]{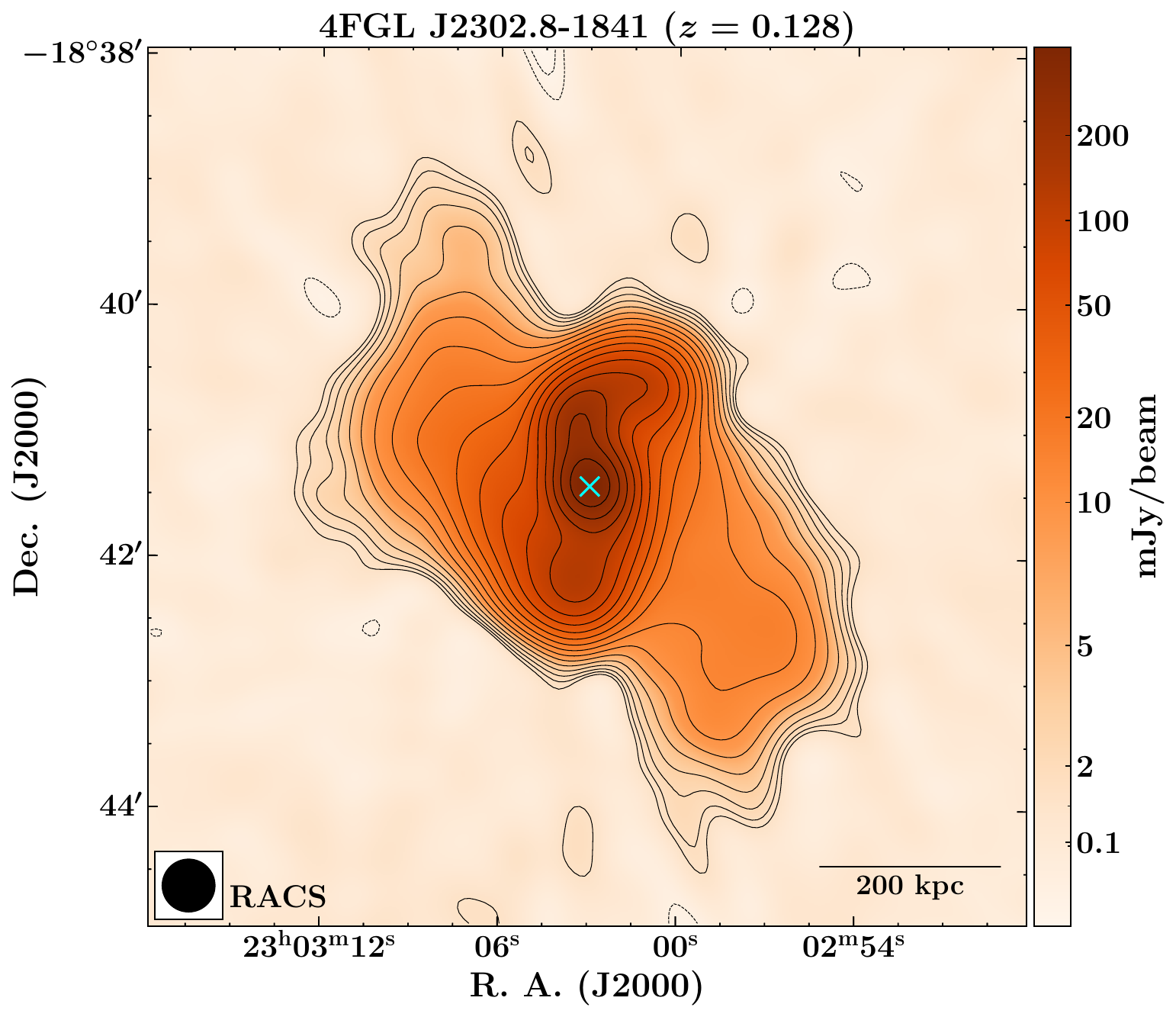}
    \includegraphics[scale=0.23]{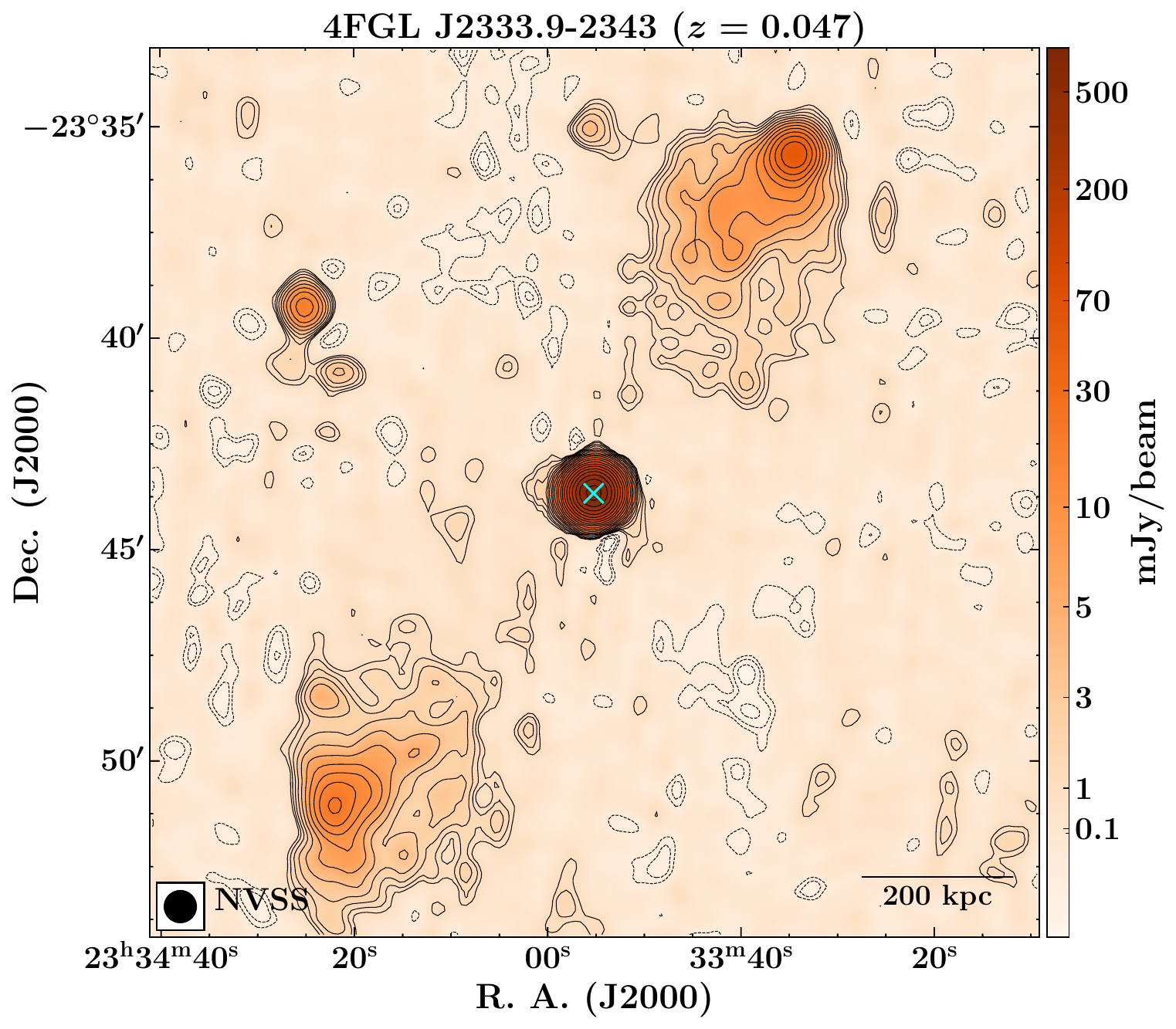}
    }
\caption{Figure 1 (continued)}
\end{figure*}

The parameters $\alpha_{\rm ext}$ and $\alpha_{\rm core}$ are spectral indices {($F(\nu) \propto \nu^\alpha$)} of the extended and core emission, respectively, and were taken as $\alpha_{\rm ext}=-0.8$ and $\alpha_{\rm core}=0$. The core and extended flux densities, $F_{\rm core}$ and $F_{\rm ext}$ ($=F_{\rm total}-F_{\rm core}$), were calculated at rest-frame 3 GHz assuming the spectral indices mentioned above. We derived $F_{\rm total}$ using the survey data with the largest restoring beam to consider all the diffuse extended flux. To estimate $F_{\rm core}$, we used the high-resolution VLASS data, which allowed us to distinguish the core component from extended lobes properly. Typical uncertainty in the flux densities, including calibration errors are about 10\% \citep[see, e.g.,][]{2021PASA...38...58H,2022A&A...659A...1S}. Furthermore, we collected the overall radio spectral indices for all sources using SPECFIND (v3.0) catalog \citep[][]{2021A&A...655A..17S}. The core dominance and radio spectral indices along with $F_{\rm core}$ and $F_{\rm total}$ values are provided in Table~\ref{tab:basic_info}.

Spectroscopic redshifts are available in the literature for all sources except 4FGL J0126.5$-$1553 or PMN J0127$-$1556. We adopted the photometric redshift of $z=0.988$ for this object from \citet[][]{2022Univ....8..587F}. The optical spectra of 4 objects, 4FGL J0312.9+4119, 4FGL J2302.8$-$1841, 4FGL J2253.3+3233, and 4FGL J2333.9$-$2343, show broad emission lines suggesting them to be powered by radiatively efficient accretion process (see references in Table~\ref{tab:basic_info}). For the remaining 11 sources, absorption features from the host galaxy stellar population dominate their optical spectra. The lack of strong emission lines suggests they are likely LERG-type objects.

\begin{figure*}
\gridline{\fig{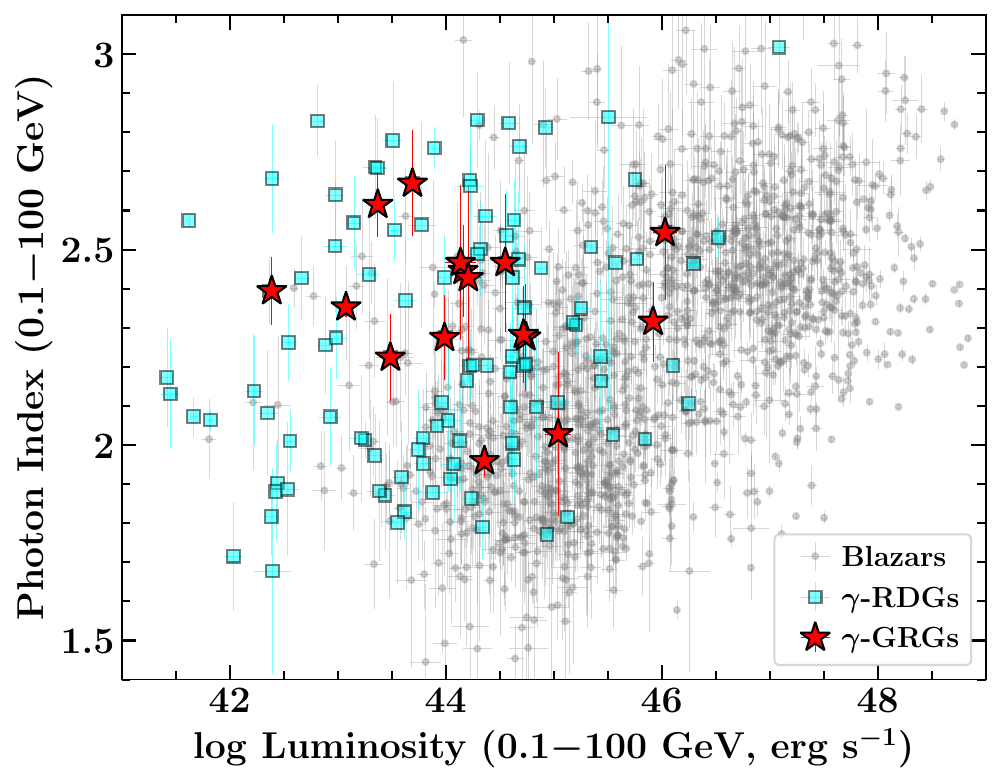}{0.47\textwidth}{(a)}
          \fig{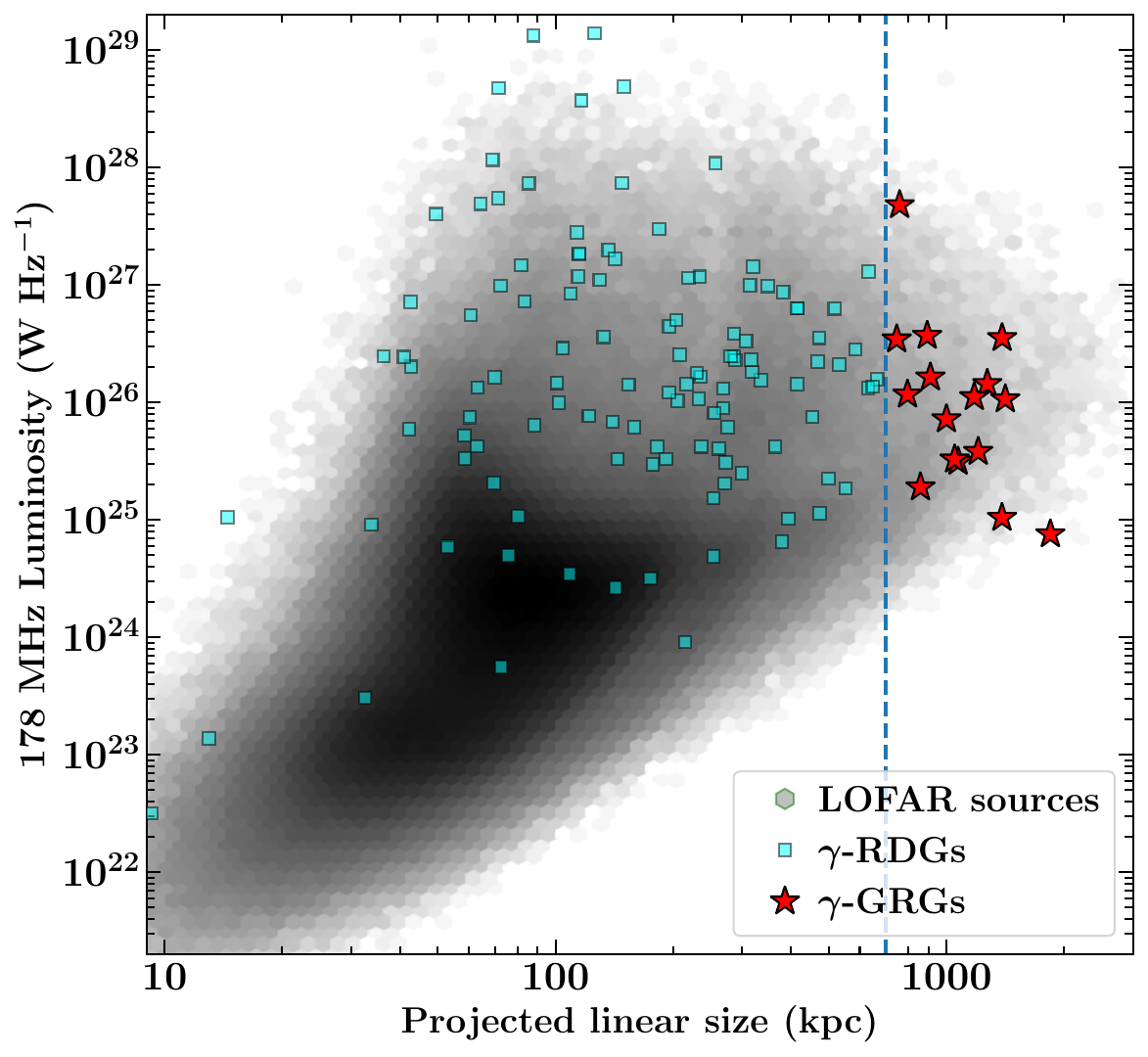}{0.4\textwidth}{(b)}
          }
\caption{Left: The \gm-ray luminosity versus photon index plane for \gm-ray emitting jetted sources. For a comparison, we also show blazars with gray circles. Right: This plot shows the variation of 178 MHz radio power with the projected linear size. The gray data points belong to LOFAR detected AGN. The vertical dashed line shows the boundary line (700 kpc) for a source to be termed as a GRG. In both panels, the red stars represent the \gm-ray detected GRGs and the cyan squares refer to the general \gm-ray emitting misaligned AGN population.}
\label{fig:cd}
\end{figure*}
\section{Discussion}\label{sec4}
Identifying GRGs in the high-energy \gm-ray band is important since they are thought to lie close to the plane of the sky. Hence, the observed radiation is not expected to be significantly Doppler-boosted. Therefore, these objects can be promising candidates to explore the origin of the \gm-ray emission. We utilized the data from the sensitive, low-resolution surveys, LOFAR, NVSS, and RACS, and carefully inspected the morphologies of all \gm-ray sources. This exercise identified 16 objects with projected source sizes exceeding 0.7 Mpc. Among them, eight are reported as GRGs for the first time in this work. These objects are highlighted in boldface in Table~\ref{tab:basic_info}. The identification of such a large number of \gm-ray emitting GRGs demonstrates the capabilities of the latest surveys in revealing the intriguing faint extended emission associated with relativistic jets. 

These surveys have also revealed a complex morphology of several of the \gm-ray emitting GRGs (Figure~\ref{fig:vlass1}). Some of them, e.g., 4FGL J0525.6$-$2008, exhibit bent radio jets resembling wide-angled tailed sources. Interestingly, $\gtrsim$50\% of the sources show diffuse, low-surface brightness extended radio emission, sometimes on one side of the source, similar to FR~Is. This finding is in contrast with general non-\gm-ray detected GRG population where the known samples are reported to be dominated by FR~II radio sources \citep[cf.][]{1999MNRAS.309..100I,2018ApJS..238....9K,2020A&A...635A...5D,2020A&A...642A.153D}.

GRGs usually show a steep radio spectrum \citep[$\alpha\sim-0.75$;][]{2020A&A...635A...5D,2020A&A...642A.153D}. Interestingly, most of the \gm-ray emitting GRGs exhibit a flat or inverted radio spectrum ($\alpha\gtrsim-0.5$). The radio spectral index was found to be $<-0.5$ only for four sources (Table~\ref{tab:basic_info}). Since these objects are \gm-ray emitters, they might be viewed at a relatively small viewing angle compared to the general, non-\gm-ray detected GRG population, leading to the observation of a flatter radio spectrum.

The core dominance is found to be small ($<$1) for \gm-ray detected radio galaxies \citep[][Paper I]{2019A&A...627A.148A}. On the other hand, blazars exhibit a core-dominated radio emission \citep[$C_{\rm r}D>$1;][]{2001MNRAS.326.1455M,2015MNRAS.451.4193C}. For four sources, we found $C_{\rm r}D>0.5$, indicating a possible beamed core emission. Indeed, one of them, 4FGL J2333.9$-$2343 or PKS 2331$-$240, has been reported to host a blazar nucleus, suggesting a possible jet bending with the inner jet aligned towards us \citep[][]{2017A&A...603A.131H,2023MNRAS.525.2187H}. Another source, 4FGL~J0946.0+4735 or RX J0946.0+4735, was earlier classified as a BL Lac object. Moreover, there are four \gm-ray emitting GRGs with $C_{\rm r}D>0$ (on log-scale), indicating a non-negligible core contribution. This finding is consistent with the relatively flat radio spectrum observed from some of these objects. The lobe emission was found to be dominating over the core for the remaining GRGs suggesting a larger viewing angle of the jet. The observed \gm-ray emission from these objects could have significant contribution from the lobes similar to that observed from Cen A and NGC 6251 \citep[cf.][]{2009ApJ...695L..40A,2019MNRAS.490.1489P,2024ApJ...965..163Y}. 

In the left panel of Figure~\ref{fig:cd}, we show the distribution of the \gm-ray photon index with the \gm-ray luminosity for Fermi-LAT detected AGN. As can be seen, the distribution of the \gm-ray detected GRGs is similar to that of the non-GRG misaligned AGN population \citep[see also,][Paper I]{2022MNRAS.513..886B}. This result hints at the similar radiative processes working for both source populations and radiating the \gm-ray emission. Furthermore, most GRGs have soft \gm-ray spectra (photon index $>$2), indicating the high-energy radiation to be produced by the tail-end of the electron population. This is because the \gm-ray softening due to extragalactic background light attenuation may not be important since all sources are located in the nearby Universe \citep[$z<1$; cf.][]{2015ApJ...813L..34D}. Since most of the sources have only a weak accretion activity, as revealed by the lack of strong optical emission lines, the \gm-ray absorption due to pair production with the broad line region photon field is also not expected to be crucial \citep[cf.][]{2016ApJ...821..102B}. A detailed multiwavelength study of these objects may provide clues about the origin of the \gm-ray emission.

We used the LOFAR results to plot the power/linear-size ($P-D$) diagram and examine the location of \gm-ray emitting GRGs in it \citep[Figure~\ref{fig:cd}, right panel; e.g.,][]{1982IAUS...97...21B,1997MNRAS.292..723K,2019A&A...622A..12H,2023A&A...678A.151H}. Assuming a spectral slope of $-$0.8, we extrapolated the observed total flux densities of the \gm-ray emitting misaligned AGN, including GRGs, to that at 178 MHz in the rest frame. The distribution of the \gm-ray detected objects was similar to that of the genral LOFAR-detected AGN population.

\section{Summary}\label{sec6}
We have utilized the high-quality, low-resolution radio survey datasets to search for GRGs in the sample of astrophysical objects detected with the Fermi-LAT. We summarize our results as follows:

\begin{enumerate}
    \item We report the identification of 16 GRGs among the \gm-ray sources included in the 4FGL-DR4 catalog, including eight identified as GRGs for the first time.
    \item Several \gm-ray detected GRGs exhibit complex radio morphologies though a significant fraction ($\gtrsim$50\%) show low surface brightness features characteristic of FR I radio sources. This finding is in tension with non-\gm-ray emitting GRGs, which have been reported to be dominated by FR~II-shaped morphologies.
    \item We derived the core dominance at rest-frame 3 GHz and found several objects to have core-dominated emission ($C_{\rm r}D>0$, on log-scale). On the other hand, the non-\gm-ray detected GRGs usually have lobe-dominated emission, as reported in earlier studies. The differences could be because the \gm-ray detection indicates a relatively small viewing angle of the jet, which can explain their significant core dominance.
    \item In the \gm-ray luminosity versus \gm-ray photon index plane and $P-D$ diagram, \gm-ray detected GRGs tend to occupy the same region as non-GRG \gm-ray emitting misaligned AGN. These results hint that the origin of the \gm-ray emission could be the same in both source populations.
\end{enumerate}

\acknowledgements

We thank the journal referee for constructive criticism.
A.D. is thankful for the support of the Proyecto PID2021-126536OA-I00 funded by MCIN / AEI / 10.13039/501100011033. G.B. acknowledges financial support for the GRACE project, selected through the Open Space Innovation Platform (\url{https://ideas.esa.int}) as a Co-Sponsored Research Agreement and carried out under the Discovery program of and funded by the European Space Agency (agreement No. 4000142106/23/NL/MGu/my). G.B. acknowledges financial support from the Bando Ricerca Fondamentale INAF 2023 for the project: \textit{\lq\lq The GRACE project: high-energy giant radio galaxies and their duty cycle\rq\rq}. 

The National Radio Astronomy Observatory is a facility of the National Science Foundation operated under cooperative agreement by Associated Universities, Inc.

LOFAR is the Low Frequency Array designed and constructed by ASTRON. It has observing, data processing, and data storage facilities in several countries, which are owned by various parties (each with their own funding sources), and which are collectively operated by the ILT foundation under a joint scientific policy. The ILT resources have benefited from the following recent major funding sources: CNRS-INSU, Observatoire de Paris and Université d'Orléans, France; BMBF, MIWF-NRW, MPG, Germany; Science Foundation Ireland (SFI), Department of Business, Enterprise and Innovation (DBEI), Ireland; NWO, The Netherlands; The Science and Technology Facilities Council, UK; Ministry of Science and Higher Education, Poland; The Istituto Nazionale di Astrofisica (INAF), Italy.

 This paper includes archived data obtained through the CSIRO ASKAP Science Data Archive, CASDA (https://data.csiro.au). This research uses services or data provided by the Astro Data Lab, which is part of the Community Science and Data Center (CSDC) Program of NSF NOIRLab.

\bibliographystyle{aasjournal}
\bibliography{Master}

\end{document}